\documentclass[twocolumn,showpacs,preprintnumbers,amssymb,nofootinbib, aps,prd, eqsecnum ]{revtex4}
\usepackage{graphicx, booktabs, epsf, epsfig}
\usepackage{bm} 
\usepackage{amsmath}

\def\be{\begin{equation}}
\def\ee{\end{equation}}
\def\beq{\begin{eqnarray}}
\def\eeq{\end{eqnarray}}
\def \bsp{\begin{split}}
\def \ensp{ \end{split} }

\def\IL{\relax{\rm I\kern-.18em L}}

\def\f{\frac}

\begin{document}

\title{New solutions of charged regular black holes and their stability }

\author{Nami Uchikata}
\email{nami@astr.tohoku.ac.jp}
\author{Shijun Yoshida}
\email{yoshida@astr.tohoku.ac.jp}
\author{Toshifumi Futamase}
\email{tof@astr.tohoku.ac.jp}
 \affiliation{Astronomical Institute, Tohoku University, Aramaki-Aoba, Aoba-ku, Sendai
980-8578, Japan}

\date{\today}

\begin{abstract}
We construct new regular black hole solutions by matching the de Sitter solution and the Reissner-Nordstr\"om solution with a timelike thin shell. 
The thin shell is assumed to have mass but no pressure and obeys an equation of motion derived from Israel's junction conditions. 
By investigating the equation of motion for the shell, we obtain stationary solutions of charged regular black holes and examine stability of the solutions. 
 Stationary solutions are found in limited ranges of  $0.87L \leq m \leq 1.99L$, and they are stable against small radial displacement of the shell with fixed values of $m,M,$ and $Q$ if $M>0$, where $L$ is the de Sitter horizon radius, $m$ the black hole mass, $M$ the proper mass of the shell and $Q$ the black hole charge. 
All the solutions obtained are highly charged in the sense of $Q/m >2 \sqrt{3} \approx 0.866$. 
By taking the massless limit of the shell in the present regular black hole solutions,  we obtain the charged regular black hole with a massless shell obtained by Lemos and Zanchin and investigate stability of the solutions. 
It is found that Lemos and Zanchin's regular black hole solutions given by the massless limit of the present regular black hole solutions permit stable solutions, which are obtained by the limit of $M \to 0 $.
\end{abstract}

\pacs{04.70.-s}

\maketitle
\newpage
\section{Introduction}
One of the most interesting questions in general relativity concerns the inner structure of black holes. 
However, it is hard to give a definite answer to this question because of the existence of a spacetime singularity 
where the curvature diverges indefinitely and general relativity breaks down. 
Penrose and Hawking proved that the gravitational collapse with physically reasonable initial conditions 
inevitably leads to the formation of a singularity, known as the singularity theorems \cite{pen, haw, he}.
Although such singularities are supposed to be concealed by the event horizon and to be isolated from the domain of predictability, 
this means that we cannot describe the entire spacetime by the present physics.
However, there may be situations where some of the assumptions in the singularity theorem may not be applied, such as the 
existence of the cosmological constant somewhere in the spacetime region. 
Thus, it is of interest if we can construct some models of black holes without spacetime singularities. 

Black holes having the regular centers are called regular black holes or nonsingular black holes. 
Existing regular black hole solutions may be divided into two classes. 
Solutions belonging to one class are characterized by the property that the black hole spacetime is sufficiently smooth everywhere. Bardeen gave this type of a solution for the first time \cite{bar}. 
The metric of his solution asymptotically approaches  the de Sitter and the Reissner-Nordstr\"om solutions  in the limit of $r \to 0$ and $r \to \infty$, respectively, where $r$ is a Schwarzschild-type radial coordinate. 
If one chooses appropriate parameters for the Bardeen solution, 
its spacetime has two horizons and looks like a Reissner-Nordstr\"om black hole with a regular center. 
Although it was thought that the Bardeen solution cannot be an exact solution of Einstein equations, Ay\'on-Beato and Garcia \cite{ab2}  showed
that  the Bardeen solution is given as a gravitational field coupled to a nonlinear magnetic monopole. 
So far, there have been many investigations based on Bardeen's work for uncharged cases \cite{dym, dym2, an} and for charged cases \cite{ab, ab2, ab3, more}. 
The other class of regular black hole solutions is composed of  the solutions constructed by matching two distinct spacetimes with a thin transition layer or surface. 
Typical solutions of this class are composed of a single regular de Sitter core  and exterior black hole spacetime between which a single thin shell exists 
\cite{fmm, lake, lemos}. 
The layer, which must be located within the event horizon because we consider the regular black hole, can be either a spacelike \cite{fmm, bal}, 
timelike \cite{lemos} or null hypersurface \cite{lemos}. 
The regular uncharged spherically symmetric black holes \cite{fmm} are motivated by the assumption that the spacetime curvature has an upper limit which is of the order of the Planck scale and the quantum effects become dominant so that the formation of the singularity is avoided. The collapsing matter will turn into a de Sitter phase when the curvature approaches a critical value. This idea was first suggested for the cosmological context by Sakharov \cite{sa} and Gliner \cite{gl}. 

As mentioned before, studies on the regular black holes are closely related to matching problems of two different spacetimes and motion of the thin shell. 
Israel derived convenient matching conditions of spacetimes for non-null transition layers~\cite{israel}. 
Barrab\'es and Israel generalized  Israel's junction conditions to a unified description including null hypersurface cases \cite{ba}. 
(For the spherically symmetric cases, see Ref. \cite{fay}.)  
As for motion of the shell, gravitational collapse of a charged shell has been studied in Refs.~\cite{chase, dlc, boul, ku}. 
(For the cases of higher dimensional spacetimes and other gravitational theory, see, e.g., Refs. \cite{gao, dgl}. )
In those studies, the interior spacetime of the shell is assumed to be flat \cite{boul, ku} or the Reissner-Nordstr\"om solution 
with different mass and charge from outside one \cite{chase}. 
As argued in those studies, the shell can be stationary and stable only if the shell has pressure. 
In cases of no pressure, the shell keeps on collapsing or expanding, or the shell collapsing (expanding) at the beginning will turn to expand (collapse). 

Stability of regular black holes is also important because unstable solutions cannot occur in nature. 
For the regular black holes with shells, instability of the stationary shell immediately implies instability of the regular black hole. 
Balbinot and Poisson \cite{bal} have analyzed stability of spacelike shells of the uncharged spherically symmetric regular black holes considered by Frolov {\it et al.} \cite{fmm}. 
They showed that for a certain parameter, the shell can be stationary and stable. 
In this study, we will apply Balbinot and Poisson's method to the cases of charged regular black holes; the regular black holes consist of a timelike massive charged shell which separates the de Sitter and the Reissner-Nordstr\"om spacetime inside the inner horizon of the Reissner-Nordstr\"om solution. 
Lemos and Zanchin have considered this type of charged regular black holes and obtained exact solutions assuming the shell is massless and pressureless \cite{lemos}. 

Our aim in this study is threefold; 
to find new regular black hole solutions having a massive thin shell, to examine their stability, and 
to examine stability of Lemos and Zanchin's regular black holes. 
For simplicity, we assume the shell is constructed of dust, i.e., the shell has mass but no pressure. 
Although the shell is pressureless in this study, we consider the de Sitter spacetime inside the shell, i.e., 
there exists matter that corresponds to a cosmological constant inside the shell. Thus, the pressureless shell can be in stationary states. 
Balbinot and Poisson \cite{bal} considered that the de Sitter horizon is of the order of the Planck scale and it is much smaller than the event horizon. 
However, we do not {\it a priori} make any assumptions for physical scales of the parameters regarding the regular black hole in this study. 

The plan of this paper is the following. In Sec. II, we briefly describe formalism for a thin shell using the 3+1 decomposition of 
Einstein equations and derive equations of motion for a thin dust shell. In Sec. III, we show results of new regular black hole solutions and their stability.
Stability of Lemos and Zanchin's regular black holes is argued in Sec. IV. Then, the conclusion is in the last section.  
Throughout this paper, we use the units of $c=G=1$, where $c$ and $G$ are the speed of light and the gravitational constant, respectively.
\section{Formulation} 
\subsection{Preliminary}
As mentioned before, we consider solutions of Einstein equations in which two different exact solutions, the de Sitter and Reissner-Nordstr\"om solutions, 
are matching by a massive thin shell. Following Ref.\cite{bal}, in this subsection, we concisely describe the formalism treating motion of the thin shell 
sandwiched between two arbitrary solutions. 

Let $V^{\pm}$ be the four dimensional spacetimes that have metrics $g_{\alpha \beta} ^{\pm}$ and a system of coordinates ${x_ \pm }^{\alpha}$. 
Let $\Sigma$ be a hypersurface described by intrinsic coordinates $\xi^a =(\xi^1,\xi^2,\xi^3)$ and located at the boundaries of $V^+$ and $V^-$. Here and henceforth, 
we use the greek and the roman lowercase letters to describe indices of the four-dimensional spacetime and  of the three-dimensional hypersurface, 
respectively. 
Let $n^{\alpha}$ be a unit normal vector to the hypersurface $\Sigma$. Thus, $n^\alpha$ has to satisfy 
\be
n^{\alpha} n_{\alpha} =\epsilon, \quad e^{\alpha} _a n_{\alpha}=0,
\ee
where $e^{\alpha} _a$ is the basis vector on $\Sigma$, defined by 
\be
e^{\alpha} _a =\f {\partial x^{\alpha}} {\partial \xi^a}\,. 
\ee
Here, $\epsilon =1$ ($\epsilon=-1$) when the hypersurface is timelike (spacelike).
The induced metric $h_{ab}$ and the extrinsic curvature $K_{ab}$ associated with $\Sigma$ are, respectively,  defined by   
\be
h_{ab} \equiv g _{\alpha \beta} e^{\alpha} _a e ^{\beta} _b\,, \quad K_{ab} \equiv  - n _{\alpha | \beta} e^{\alpha} _a e ^{\beta} _b.
\ee
Here and henceforth, we denote the covariant differentiation associated with $g_{\alpha \beta}$ and $h_{a b}$  by the stroke $(|)$ and the semicolon $(;)$, respectively. 
To describe motion of the three-dimensional hypersurface, it is useful to rewrite the basic equations in the three-dimensional form. These equations can be derived by contracting the four-dimensional quantities by $n^{\alpha}$ and/or $e^{\alpha} _a$. By using the Einstein tensor $G_{\alpha\beta}$ contracted by $n^{\alpha}$ and/or 
$e^{\alpha} _a$, and the Gauss-Codazzi equations, we obtain 
\be
\begin{split}
& -2 \epsilon G_{\alpha \beta} n^{\alpha} n^{\beta} = {}^3R +\epsilon(K^{a b} K_{ab} -K^2), \\
& G_ {\alpha \beta} e^{\alpha} _a n^{\beta} = K_{; a} -K^b _{\, \, a;b}\,, 
\end{split}
\label{gc}
\ee
where $^3 R$ is the three-dimensional Ricci scalar associated with $h_{ab}$ and $K= h_{ab } K^{ab}$.
The energy-momentum tensor on the hypersurface, $S_{ab}$, is given by a jump of the extrinsic curvature on $\Sigma$ (see, e.g., Ref.~\cite{israel}),
\be
8 \pi S_{ab} =\epsilon ([ K_{ab} ] -h_{ab} [K]) ,
\label{sab}
\ee
where $[K_{ab}] =K_{ab} | ^+ -K_{ab} | ^-$. $K_{ab} | ^{\pm}$ is evaluated on $\Sigma$ by  taking limit from $V^{\pm}$. 
Then, the energy-momentum conservation equation on the hypersurface will yield
\be
S^a_{\, \, b ; a} +\epsilon [ T_{\alpha\beta} e^{\alpha} _b n^{\beta}] = 0.
\label{dsa}
\ee

So far, we have shown  the energy-momentum conservation equation in the general form. 
We next show how these equations are given for the dust shell. If the shell is composed of dust, the stress-energy tensor of the shell is given by 
\be
S_{ab} = \sigma u_a u_b,
\ee
where $\sigma$ is the surface energy density of the shell  and $u^a$ is the matter velocity on $\Sigma$ if the shell is a timelike hypersurface. 
Thus, the energy-momentum conservation \eqref{dsa} leads to 
\be
(\sigma u^a) _{; a} = [T_{\alpha\beta } u^{\alpha} n^{\beta}]  ,
\label{ene}
\ee
where $u^{\alpha }$ is the four-velocity of the shell given by $u^{\alpha} = u^a e^{\alpha} _a$. 
 The transverse acceleration of the shell is expressed by 
\be
a^{\alpha} \equiv u^{\alpha} _{\, \, | \beta} u^{\beta}= u^a_ {\,  \, ; b} u^b e^{\alpha} _a +\epsilon u^a u^bK_{a b }n^{\alpha}. 
\ee
We are interested in the normal component of $a^{\alpha}$, which describes the motion of the shell. It is straightforward to see 
that $[n_{\alpha} a^{\alpha}] = u^a u^b [K_{ab}]$. Equation \eqref{sab} is equivalent to  
\be
[K_{ab}] = 8 \pi \epsilon \left (S_{ab} - \f {S} {2} h_{ab} \right ), \quad \mbox{ with} \quad S=h_{ab} S^{ab}\,. 
\label{ktos}
\ee
Then, we have an equation of motion of the shell, 
 \be
   n_{\alpha} a^{\alpha}| ^{+} - n_{\alpha} a^{\alpha}| ^{-} = 4 \pi \epsilon \sigma\,. 
\label{eom1}
\ee
We may also construct an equation from the arithmetic means of the extrinsic curvatures, $(K^+_{ab} + K^-_{ab})/2$. However, it is not a useful equation for the present situation. 

\subsection{Equation of motion of the shell}
In order to have a regular center by matching the de Sitter and the black hole spacetimes with a massive thin shell, at least two horizons including extremal cases, in which two horizons coincide, are required \cite{bro}. 
For uncharged spherically symmetric cases, since the Schwarzschild black hole, which is the outer solution for this situation, has a single horizon (event horizon),  the second horizon must be the de Sitter one. 
This implies that the shell has to be located between the outer event horizon and the inner de Sitter horizon and that it necessarily has to be spacelike. 
For charged and/or rotating cases, the black hole solution has double horizons, the event and Cauchy (inner) horizons. 
We may, therefore, choose any type of shell---timelike, spacelike, or null. 
Since it is physically natural to assume that the shell is a timelike hypersurface, we choose the shell to be located inside the inner horizon of the black hole solution in these cases.

Assuming $V^+$ and $V^-$ to be the Reissner-Nordstr\"om and de Sitter spacetimes, respectively, we apply the formalism given in the previous subsection to the present situation. 
We derive the equation of motion for the shell, given by $r=R(\tau)$, from Eqs.\eqref{ene} and \eqref{eom1}, with $\tau$ being the proper time of the shell. 
In this study, as mentioned, we assume the shell to be a timelike hypersurface. 
Thus, the shell radius $R$ is assumed to satisfy $R(\tau) < r_-$ and $R(\tau)<L$, where $r_-$ and $L$ denote 
radii of the inner horizon of the Reissner-Nordstr\"om solution and the de Sitter horizon, respectively. 
The spherically symmetric metric that expresses the inside and outside of the shell is written by 
\begin{equation}
 ds^2= - f(r) dt^2 +\f {1} { f(r)} dr^2 +r^2 (d \theta ^2 + \sin ^2 \theta \, d \phi^2), 
 \label{metric}
\end{equation}
and the function $f(r)$ is given by 
\be
f(r)=\left\{ 
\begin{split}
& f_{dS} (r) \equiv 1- \f {r^2} {L^2}, \quad  \mbox{for } r<R(\tau), \\
& f_{RN} (r) \equiv1 - \f {2m} {r}  +\f {Q^2} { r^2} , \quad \mbox{for }r>R(\tau),
\end{split}
\right.
\ee
where $m$ and $Q$ are the mass and the charge of the black hole, respectively. 
Here and henceforth, $H_{dS}$ and $H_{RN}$ mean functions $H$ evaluated by the de Sitter and the Reissner-Nordstr\"om solutions on $\Sigma$, respectively.
Due to the assumptions, $R(\tau) < r_-$ and $R(\tau)<L$, $f_{dS}(r)$ is always positive and $f_{RN} (r)$ has two roots (one root) for $Q<m$ ($Q=m$).
The electric potential is $\displaystyle{A^+_{\mu} d{x_+}^{\mu}=-{Q\over r} dt}$ and $A^-_{\mu} d{x_-}^{\mu}=0$. 
The four velocity of the shell is $u^{\alpha} = d x^{\alpha}/ d\tau \equiv (\dot{t}, \dot{R}, 0, 0)$, where the dot denotes the derivation with respect to $\tau$. 
Because of $u^\alpha u_\alpha = -1$, $\dot{t}$ may be written as 
\be
\dot{t} =\f {\sqrt{\dot{R}^2 + f(r)}} {f(r)} \equiv \f {\beta} {f(r)}\,,
\ee
where we have set that $t$ increases with $\tau$. The normal vector to the shell is, from $u^{\alpha} n_{\alpha}=0$ and $n^{\alpha} n_{\alpha}=1$, 
given by 
\be
n^{\alpha} =\left ( \f {\dot{ R}} {f(r)}, \beta, 0,0 \right ),
\ee
where we have considered the normal vector pointing from the de Sitter spacetime to the Reissner-Nordstr\"om spacetime. 
The induced metric on the shell is 
\be
(ds^2)_{\Sigma} = -d\tau^2 +R^2(\tau)  (d \theta ^2 + \sin ^2 \theta \, d \phi^2).
\ee
Since the four-velocity $u^\alpha$ and the normal vector $n^\alpha$ do not have $\theta$- and $\phi$-components, we only need to 
consider $t$ and $r$ components of the $T_{\alpha\beta}$, which leads to
\be
\begin{split}
& (T_{\alpha\beta} )_{dS}=- \f{3 g_{\alpha \beta }} {8\pi L^2},\\
& (T_{\alpha\beta} )_{RN}=- \f{ g_{\alpha \beta }} {8\pi } (\partial_r A_t)^2,\quad \mbox{for } \alpha, \beta =0,1.
\end{split}
\ee
Then, the energy conservation equation, 
\be
(\sigma u^a) _{; a} = [T_{\alpha\beta } u^{\alpha} n^{\beta}] =0\,, 
\ee
leads to
\be
(R^2 \sigma )^ . =0.
\ee
Note that $T_{\alpha\beta } u^{\alpha} n^{\beta}$ vanishes when $T_{\alpha\beta } \propto g_{\alpha \beta}$.
If we define the proper mass of the shell by $M= 4 \pi R^2 \sigma$, the above equation means $M$ is independent of $\tau$. 
Nonvanishing components of the extrinsic curvature are given by 
\be
n_{\alpha} a^{\alpha} =  K_{ \tau \tau}=  \f {\dot{\beta}} {\dot{R}}, \quad K^{\theta}_{\theta}=K^{\phi}_{\phi}=  - \f {\beta} {R}\,. 
\label{kab}
\ee
From  Eq.\eqref{eom1}, thus, we have 
\be
 \dot{\beta} _{RN} -  \dot{\beta} _{dS} = 4 \pi\dot{R} \sigma =-4 \pi(R \sigma)^.  ,
 \label{bdot}
\ee
where we have used $(R^2 \sigma) ^. =0$.  Integrating Eq.\eqref{bdot}, we obtain
\be
 \sqrt{\dot{R}^2 +1 - \f {R^2} {L^2}}-  \sqrt{\dot{R}^2 + 1- \f{2 m} {R} + \f {Q^2} {R^2}} = \f {M} {R} +C ,
 \label{eom2}
\ee
where $C$ is an integration constant. 
From Eqs. \eqref{sab} and \eqref{kab}, we have,    
$- (\beta_{RN} -\beta_{dS}) /R =[K^{\theta} _{\, \, \theta }] = [K^{\phi}_{ \, \, \phi}] = 4 \pi \sigma$. 
Thus, the integration constant $C$ appearing in Eq. \eqref{eom2} has to be zero.  
Rather than using Eq.\eqref{eom2}, we employ a more convenient form, 
\be
 \dot{R}^2+V(R)=-1\,, 
 \label{eom3}
\ee
where 
\be
V(R)= - \left ( \f {\f {R^3} {L^2} + \f {Q^2} {R} - 2m} {2M} - \f { M} {2R} \right )^2 -\f {R^2} {L^2 } \,. 
  \label{effpot}
\ee
Equation  \eqref{eom3} is a kind of energy conservation law because one may interpret $V(R)$ as an effective potential. 
This equation is nothing but an equation of motion for the massive thin shell located on the surface of the de Sitter sphere.  
The stationary solutions, $R={\rm const.}$,  can be obtained by solving $V(R)=-1$ and $dV(R) /dR =0$ simultaneously. 
The stability of the stationary solutions can be checked by the condition $d ^2 V(R) /d R^2 >0$ at the stationary point, i.e., 
whether or not the stationary shell is at a local minimum of the effective potential $V(R)$. From Eq. \eqref{effpot}, we see 
$V=V(|M|)$. To obtain values of $M$, we use Eq. \eqref{eom2} after obtaining solutions of the regular black hole. 
Since, in this study, we are concerned with the regular black hole solution, we assume that no naked singularity occurs, i.e., $m >  Q$.

\section{results}
In order to show numerical results, we employ the units of $L=1$, e.g., $R/L \to R$, $M/L \to M$, $m/ L \to m$, and $Q/L \to Q$. 
Let us describe a method to obtain the equilibrium states of the regular black hole. 
To obtain stationary solutions numerically, we solve $V(R)+1=0$ and $dV(R)/dR=0$ simultaneously with a Newton-Raphson--like iterative scheme. 
During the iteration procedure, values of $|M|$ and $m$ are kept constant. 
Thus, the two algebraic equations, $V(R)+1=0$ and $dV(R)/dR=0$, contain only two unknown parameters, $R$ and $Q$. 
Then, we can obtain a regular black hole solution if the iteration procedure successfully ends.
After obtaining solutions, we obtain values of $M$ from Eq. \eqref{eom2} and check their sign of $d^2V(R)/dR^2$ to see their stability. 
If $d^2V(R)/dR^2 >0$ ($d^2V(R)/dR^2 <0$), then the solution is stable (unstable).
Since, as mentioned before,we  assume that there is no naked singularity, and the shell is inside the inner horizon of 
the black hole solution, we are concerned with the solutions satisfying conditions $m >Q$, $R<1$, and $R<r_-$. 
Otherwise, we do not admit the solutions as those of the regular black hole model that we consider in this study. 
 Although the solutions of $M<0$ are not physically acceptable in normal situations, they are allowed in Eq. \eqref{eom3} and might be useful in some exotic situations. 
Thus, we show the results of the $M<0$ case as well in this study (see, e.g., Ref.\cite{boul}). 

It is helpful to examine properties of the effective potential $V$ in order to understand how the regular black hole solution is obtained.  
In Figs. 1 and 2 and Figs. 3 and 4, we show typical behaviors of the potential as functions of $R$ for stable and unstable stationary solutions, respectively.  
Figures 2 and 4 are magnified figures of Figs. 1 and 3 around extremal points, respectively. 
The potential of the stable (unstable) configuration, given in Figs. 1 and 2 (Figs. 3 and 4), are characterized by a set of 
parameters $(m,Q,|M|)=(1.2, 1.191, 0.132)$ [$(m,Q,|M|)=(1.31, 1.309, 0.01)$]. 
A local minimum (maximum) of the potential, shown in Figs. 1 and 2 (Figs. 3 and 4), is at $R=0.87661$ ($R=0.86623$). 
As can be seen from Eq. \eqref{effpot}, the effective potential $V$ diverges to minus infinity as $R \to 0$ and as $R \to \infty$, which means that $V$ has at least one maximum. 
This property may partly be confirmed in Figs. 1--4. 
It is also observed in Figs. 2 and 4 that $V \approx -1$ at the extremal points. 
Thus, one sees that these potentials permit the stationary solutions.
 
To investigate basic properties of the regular black hole solutions, we calculate many sequences of stationary solutions. 
The sequences of stationary solutions are specified by a fixed parameter $m$, and are obtained by increasing 
a parameter $|M |$ from a minimum value $10^{-4}$ (see, also, the first paragraph of this section). 
In other words, they are given as a set of functions $R(M)$ and $Q(M)$ with a fixed parameter $m$. 
The sequences of stationary regular black hole solutions, given as functions $R(M)$ and $Q(M)$, are shown in Figs. 5 and 6. 
As shown in these figures, we obtain the regular black hole solutions for positive and negative values of $M$. 
One of the interesting findings in this study is that all the positive (negative) $M$ solutions are stable (unstable), and it seems that these stable and unstable solutions belong to continuous sequences of the stationary solutions of a fixed $m$ as can be seen in Figs. 5 and 6.  
For the sequences of the solutions obtained in this study, $R$'s are  increasing functions of $M$ (see Fig. 5), and (Q/m)'s are  decreasing functions of $M$ (see Fig. 6). 
For high $m$ sequences of the solutions, values of $R$ approach unity (see Fig. 5). 
It is important to point out that all the solutions obtained in this study satisfy $1>Q/m > 0.86$, i.e., they are highly charged, and 
that  some solutions are nearly extreme in the sense $Q \sim m$ at one end of the sequences of the solution (see Fig. 6). 

As shown in our numerical results argued so far, the regular black hole solutions are only found in some restricted parameter regions. 
Thus, it is useful to show clearly in which parameter regions the regular black hole solutions with a timelike thin shell occur.  
In Figs. 7 and 8, we show two-dimensional parameter regions where regular black hole solutions are found in this study. 
Stable and unstable solutions are found in $0.87  \leq m \leq 1.99$ with positive $M$ and in $1.3 \leq m$ with negative $M$, respectively (see Fig. 7). 
As can be observed in Fig. 7, the maximum value of $M$ $(\approx 0.43)$ for the solutions obtained in this study is achieved by 
the solution with the minimum value of $m$ $(\approx 0.87)$. 
From Fig. 8,  it is observed that the radius of the shell is in the range $0.866 < R <1$ for the solutions obtained.
In the present study, we find no upper limit of $m$ for existing unstable regular black hole solutions. 
Note that we show results of unstable solutions only for the range of $1.3 \leq m\leq 1.99$ in Fig. 7, but $m=1.99$ does not mean the upper limit.
For the cases of $m \geq 2$, we cannot determine the minimum value of $|M|$ for unstable solutions because of numerical difficulties. 

\section{discussions}
As mentioned in the Introduction, one of our aims in this study is to analyze the stability of the charged regular black holes 
with a massless thin shell derived by Lemos and Zanchin \cite{lemos}. In order to investigate regular black holes 
with a massless thin shell,  we may consider the limit of $M \to 0$ in our formalism. 
We focus on the $M>0$ case in this section since the $M<0$ case is not physically acceptable in usual situations. 
Besides, all the solutions with $M<0$ are unstable, so they are not feasible as the regular black hole models.
Thus, we may exclude the $M<0$ case for our physical interests. 
In the $M \to 0$ limit, the effective potential $V$ and its first derivative $dV/dR$ are approximated by 

\begin{eqnarray}
V&\approx& -{\left(Q^2-2 m R+{R^4\over L^2}\right)^2\over 4 M^2 R^2} \,, \\
{dV\over dR}&\approx& {\left(Q^2-3 {R^4\over L^2}\right)\left(Q^2-2 m R+{R^4\over L^2}\right) \over 2 M^2 R^3}\,. 
\end{eqnarray}
Thus, we may obtain stationary solutions in the limit of $M \to 0$ if  the following conditions are satisfied: 
\begin{eqnarray}
&& Q^2-2 m R+{R^4\over L^2} =O(M)  \,,  \\
&& Q^2-3 {R^4\over L^2} = O(M) \,. 
\end{eqnarray}
We then assume that charged regular black hole solutions in the massless limit may be expanded in terms of $M$ as follows, 
\begin{eqnarray}
{R}^2 &=&Q L/\sqrt{3}+{1\over 3} A L M+O(M^2)\,,   \label{lemo01} \\
m&=&2 {{R}^3\over L^2} +B M+O(M^2)\, ,  \label{lemo02}
\end{eqnarray}
where $A$ and $B$ are functions independent of $M$. Substituting Eqs. \eqref{lemo01} and  \eqref{lemo02} into $V$ and $dV/dR$, we obtain 
\begin{eqnarray}
V&=&-\frac{R^2}{L^2}-\frac{(2 A R+2 B L)^2}{4 L^2}+O(M) \,, \\
{dV\over dR}&=& {2 A B L-2 R+2 A^2 R \over L^2}+O(M)\,. 
\end{eqnarray}
Then, stationary solutions in the limit of $M \to 0$ are, in terms of $R_0$ and $L$, given by
\begin{eqnarray}
Q_0&=& {\sqrt{3}\over L}{R_0}^2\,, \label{lemo1}\\
m_0&=&2 {{R_0}^3\over L^2} \label{lemo2} \,, \\
A&=& {R_0 \over \sqrt{L^2-R_0^2}} \,, \label{lemo3}  \\
B&=&{1-A^2 \over A L}R_0 \,, \label{lemo4}
\end{eqnarray}
where quantities indicated by the subscript $0$ correspond to stationary solutions in the limit of $M \to 0$. 
From the conditions for the regular black hole with a timelike thin shell, $r_-> R_0$, $L>R_0$ and $m_0>Q_0$, we obtain a constraint  for $R_0$, given by $L> R_0>{\sqrt{3}\over 2}L$.  
These massless limit solutions, given by Eqs. \eqref{lemo1} and \eqref{lemo2}, are exactly the same as those given by Lemos and Zanchin, although our notation is different form theirs.
[Lemos and Zanchin also derived the relations $R \geq Q/\sqrt{3}$ and $m\leq 2 Q/\sqrt {3}$. 
(These inequalities are derived from Eqs. \eqref{lemo1} and \eqref{lemo2} and $L> R_0>{\sqrt{3}\over 2}L$.)] 
For these massless limit solutions, the second derivative of the effective potential $d^2V/dR^2$ is approximated by 
\begin{eqnarray}
{d^2V\over dR^2}&\approx&  {12 R_0 \sqrt{L^2-R_0^2}  \over M L^3} \,, \label{ddV}
\end{eqnarray}
where the conditions for the stationary solution \eqref{lemo01}, \eqref{lemo02}, and \eqref{lemo3}--\eqref{lemo2} have been used. 
For the stationary solution, we therefore see that $d^2V/dR^2 \to \infty$ as $M \to 0$. 
This means that  in the massless limit $M \to 0$, we have stable solutions.
Note that as can be seen from Eqs. \eqref{lemo3} and \eqref{lemo4}, the massless limit solutions break down when $R_0 \to L$. 
This is because as $R_0 \to L$, the shell becomes lightlike, for which the present formalism for the timelike shell is not applicable. 

The above results show that in the limit of $M \to 0$, the solutions exist only for ${\sqrt{3}/2}(\approx 0.866025)< R/L < 1$, $3\sqrt{3}/4 (\approx1.29904) < m/L < 2$, and $\sqrt{3}/2  < Q/m <1$. 
Our numerical solutions with $M=10 ^{-4}$ will satisfy the massless condition ($M=0$) with good accuracy. 
Thus, we may regard these solutions as good approximations for exactly massless solutions. 
Some of those solutions are displayed in Table I. 
In this table, we may confirm that infinitesimal quantities $R^2-QL/\sqrt{3} $ and $m-2 R^3/L^2$ and a divergent quantity $d^2 V/d R^2$ indeed depend on $M$ as given by Eqs. \eqref{lemo01}, \eqref{lemo02}, and \eqref{ddV}. 
We see that sets of parameter $(m, M,R,Q )=(1.3, 10^{-4} ,0.86626, 1.29964)$ and $(m, M,R,Q )=(1.99, 10^{-4} , 0.99865, 1.72626)$ correspond to the lower and the upper endpoints, respectively,  in the relation of the Lemos and Zanchin solutions $(Q \leq)$ $m \leq 2Q/ \sqrt{3}$. 

These analyses show that Lemos and Zanchin's regular black hole solutions, given by the massless limit of the present regular black hole solutions, permit stable solutions. 
This conclusion, however, is not a final one because a properly massless thin shell case is excluded in the present analysis and because we only consider an example of regular black hole solutions that coincide with Lemos and Zanchin's solution in a massless thin shell limit. 
Thus, further analyses are required to draw a definite answer of whether Lemos and Zanchin's regular black hole solutions are stable or not. 

Comparing these analytic results discussed so far to numerical results given in the last section, we may guess the lower and upper limits of the physical quantities for the regular black hole model to exist for the case of  $M \neq 0$. 
Then, it may be conjectured that $m/L < 2$, $\sqrt{3}/2 < Q/m <1$ and ${\sqrt{3}/2}< R/L < 1$ for the stable regular black hole solutions (see Figs. 7 and 8). 

Finally, let us consider a physical scale of the stable regular black hole solutions we obtain in this study, which have not been specified so far. 
For the stability analysis of Schwarzschild-type regular black holes in Refs.\cite{fmm, bal}, the de Sitter horizon radius is assumed to be of the order of the Planck scale and $L \ll R< r_+$, where $r_+$ denotes the event horizon radius. Thus, $L \approx l_p$, where $l_p$ denotes the Planck length. 
On the other hand, if we take the above assumption, i.e., $L \approx l_p$, our analysis, based on a classical 
approach, fails since the present solutions satisfy $R <  l_p \approx L < r_+$. 
Since the curvature invariant of the de Sitter spacetime is $R_{\mu\nu\rho\sigma} R^{\mu\nu\rho\sigma} = 24/L^4$, if there exists 
an upper bound of the curvature and our analysis is valid, the upper bound of the curvature has to be smaller than that of the Planck scale.
However, our results show that if we assume the de Sitter horizon radius is of the order of the Planck scale, the present stable charged regular black hole solution is restricted to quantum size. 
Although the de Sitter horizon radius is assumed to be other physical scales of the vacuum phase transition, such as the grand unified theory (GUT) scale, the present stable black holes are also restricted to quantum size.

\section{Conclusion}
We have constructed new regular black hole solutions by matching the de Sitter solution and the Reissner-Nordstr\"om solution with a timelike thin shell.  
The thin shell is assumed to have mass but no pressure and obeys an equation of motion derived from Israel's junction conditions. 
By investigating this equation of motion for the shell, we obtain stationary solutions of charged regular black holes and examine stability of the solutions. 
Stationary solutions are found in limited ranges of  $0.87L \leq m \leq 1.99L$, and they are stable against small radial displacement of the shell with fixed values of $m, Q$, and $M$ if $M>0$.
All the solutions obtained are highly charged in the sense of $Q/m >2 \sqrt{3} \approx 0.866$. 
By taking the massless limit of the shell in the present regular black hole solutions,  we obtain the charged regular black hole with a massless shell obtained by Lemos and Zanchin~\cite{lemos} and investigate stability of the solutions. 
It is found that Lemos and Zanchin's regular black hole solutions permit stable solutions.
 
\section*{Acknowledgements}
N.U. is supported by the GCOE Program ``Weaving Science Web beyond 
Particle-matter Hierarchy'' at Tohoku University. This work is supported by a Grants-in-Aid for Scientific Research 
from JSPS (No. 23540282 and No. 24540245 for T. F. and S. Y., respectively.) 

\clearpage
\begin{table}
\caption{Physical quantities for the stationary solutions with $|M|=10^{-4}$. }
\begin{ruledtabular}
\begin{tabular}{l l l rrr}\hline
$R$&$Q$& $m$ & $R^2-Q L/\sqrt{3} \, $ &$m-2 R^3/L^2$  &$d^2 V/d R^2$
\\ \midrule
$0.866$ & $1.29$ & $1.30$ &  $5.78 \times 10^{-5}$ &  $-1.00 \times 10^{-4}$ &  $5.19 \times 10^4$\\
$0.908$ & $1.43$ & $1.50$ &  $7.31 \times 10^{-5}$ &  $-1.55 \times 10^{-4}$ &  $4.55 \times 10^4$\\
$0.947$ & $1.55$ & $1.70$ &  $9.86 \times 10^{-5}$ &  $-2.48 \times 10^{-4}$ &  $3.63 \times 10^4$\\
$0.998$ & $1.72$ & $1.99$ &  $6.41 \times 10^{-4}$ &  $-5.10 \times 10^{-4}$ &  $5.47 \times 10^3$\\

 \bottomrule

\end{tabular}
\end{ruledtabular}
\end{table}

\begin{figure}
\includegraphics [height =5.8cm, width=7.5cm] {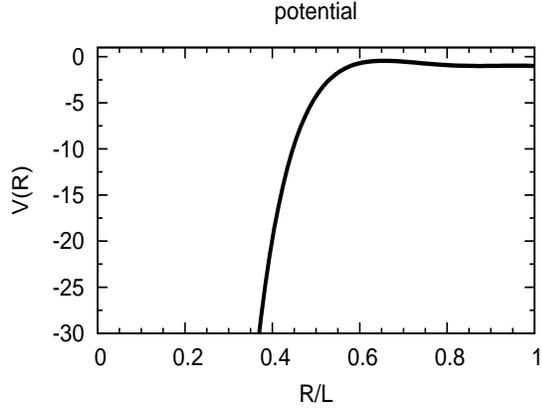}
\caption{Effective potential $V(R)$ for the stable solution with $m=1.2$, $Q=1.1913$ and $|M|=0.132$. The local minimum is at $R=0.87661$.} 
\end{figure} 
 
\begin{figure}
\includegraphics [height =5.8cm, width=7.5cm] {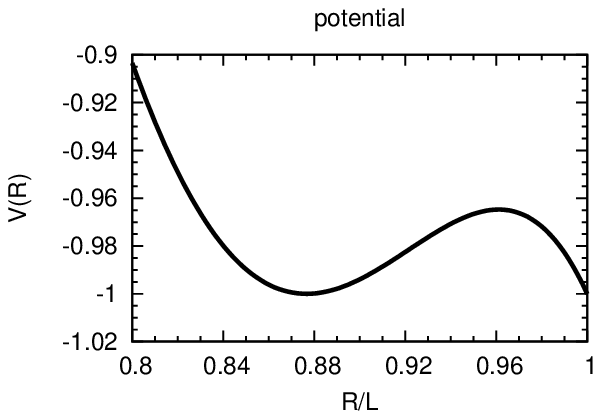}
\caption{Magnified figure of Fig. 1 around the local minimum.}

\includegraphics [height =5.8cm, width=7.5cm] {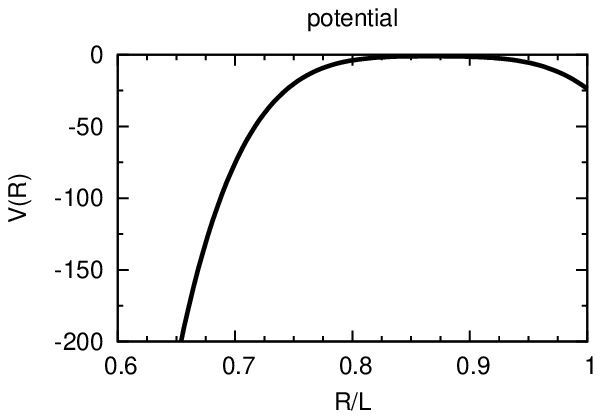}
\caption{Effective potential $V(R)$ for the unstable solution with $m=1.31$, $Q=1.3909$ and $|M|=0.01$. The local maximum is at $R=0.86623$.} 
  
\includegraphics [height =5.8cm, width=7.5cm] {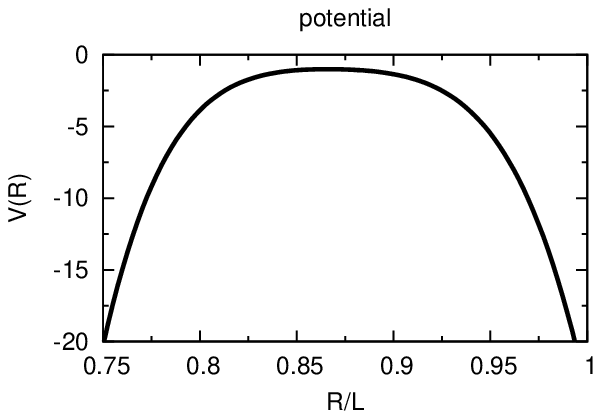}
\caption{Magnified figure of Fig. 3 around the local maximum.}
\end{figure}
  
\begin{figure} 
\includegraphics [height =5.8cm, width=7.5cm] {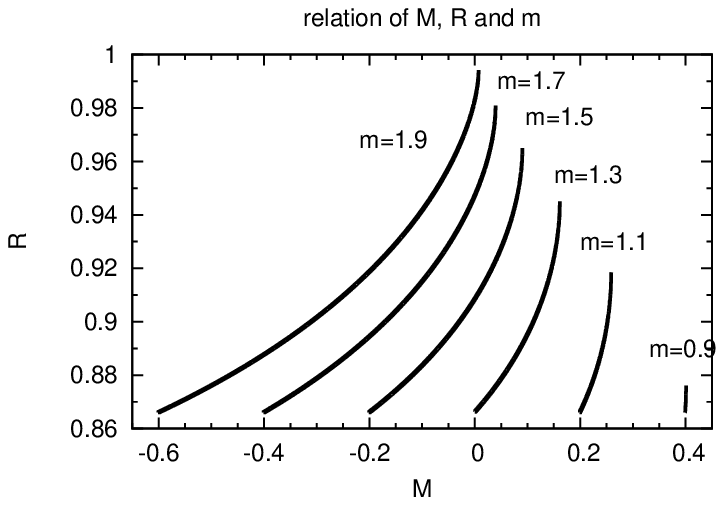}
\caption{Radius of the shell $R$ for the sequences of the stationary regular black holes with fixed $m$, given as functions of
 the mass of the shell $M$. From bottom right to top left, the curves correspond to the sequences 
with $m=0.9$, $1.1$, $1.3$, $1.5$, $1.7$ and $1.9$.}

\includegraphics [height =5.8cm, width=7.5cm] {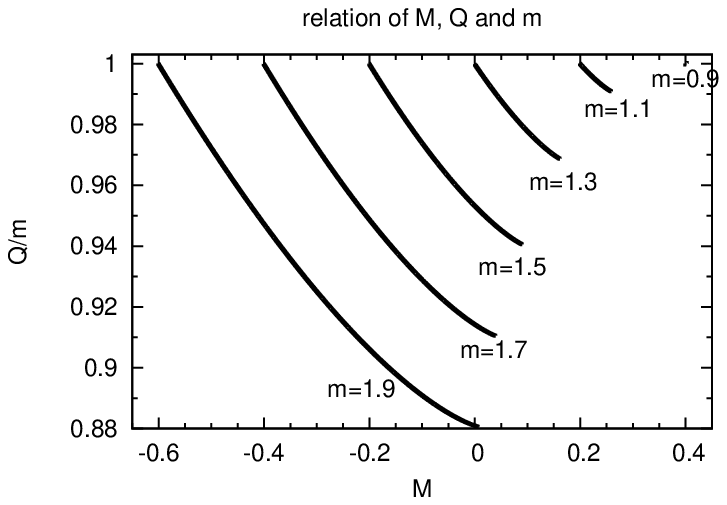}
\caption{Ratio of the black hole charge to the black hole mass $Q/m$ for the sequences of the stationary regular black hole 
with fixed $m$, given as functions of the mass of the shell $M$. From top right to bottom left, the curves correspond to 
the sequences with $m=0.9$, $1.1$, $1.3$, $1.5$, $1.7$ and $1.9$. }

\end{figure}

\begin{figure} 

\includegraphics  [height =5.8cm, width=7.8cm] {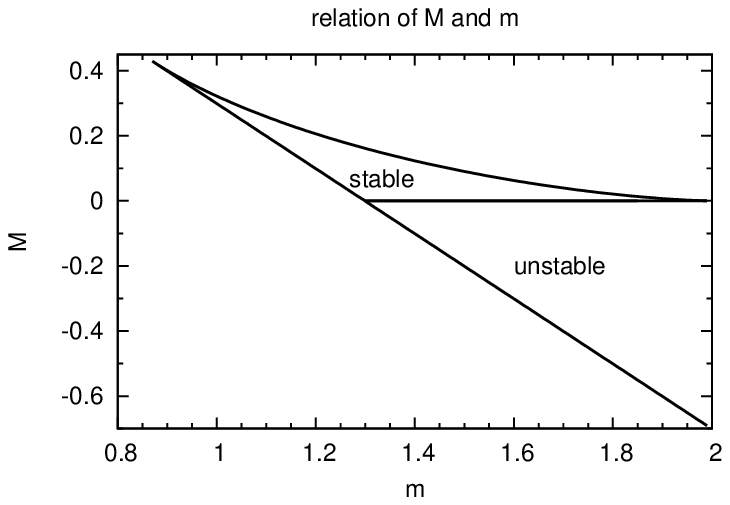}
\caption{Parameter region in $(m,M)$ plane where the regular black hole solutions are found. 
The stable (unstable) solutions exist in the positive (negative) $M$ region as indicated by the labeles. }

\includegraphics [height =5.8cm, width=7.5cm] {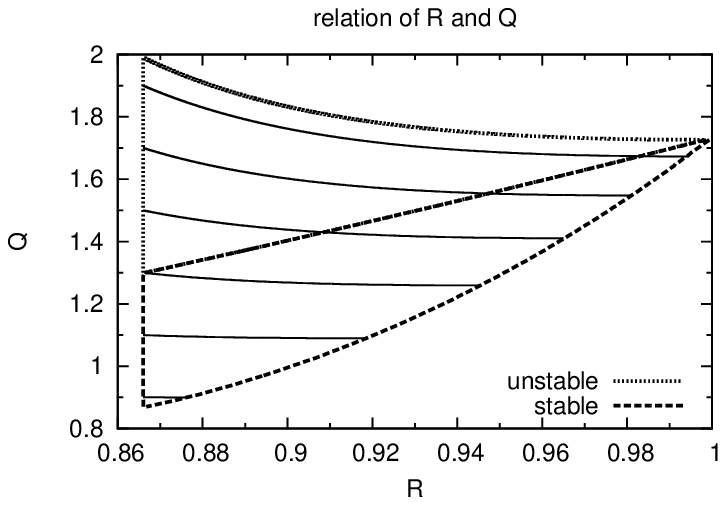}
\caption{Parameter region in $(R,Q)$ plane where the regular black hole solutions are found. 
The stable (unstable) solutions exist in the domain bounded the dotted (short dashed) curves. 
The long dashed curve indicates the boundary between the stable and unstable solution regions, 
given as the $|M|=10^{-4}$ sequence of the solutions. From bottom to top, the solid curves correspond to the sequences 
with $m=0.9$, $1.1$, $1.3$, $1.5$, $1.7$ and $m=1.9$.}

 \end{figure}
  
  \end{document}